\documentclass[useAMS,usenatbib]{mn2e}
\usepackage{aas_macros}
\usepackage{epstopdf}
\usepackage{epsfig}
\usepackage{natbib}
\usepackage{float}
\usepackage{amsmath}
\usepackage{times}
\title{The connection between the UV colour of early type galaxies and the stellar initial mass function revisited}

\author[Zaritsky, Gil de Paz, \& Bouquin]
{
Dennis Zaritsky$^{1}$, Armando Gil de Paz$^2$, and Alexandre Y. K. Bouquin$^2$\\
$^{1}$Steward Observatory, University of Arizona, 933 North Cherry Avenue, Tucson, AZ 85721, USA; dennis.zaritsky@gmail.com\\
$^{2}$Departamento de Astrof\'isica y CC$.$ de la Atm\'osfera, Universidad Complutense de Madrid, Madrid, Spain \\
}

\begin{document}

\date{}

\pagerange{\pageref{firstpage}--\pageref{lastpage}} \pubyear{2014}

\maketitle

\label{firstpage}

\begin{abstract} 
We extend our initial study of the connection between the UV colour of galaxies and  both the inferred stellar mass-to-light ratio, $\Upsilon_*$, and a mass-to-light ratio referenced to Salpeter  initial mass function (IMF) models of the same age and metallicity, $\Upsilon_*/\Upsilon_{Sal}$,
using new UV magnitude measurements for a much larger sample of early-type galaxies, ETGs,  with dynamically determined mass-to-light ratios. We confirm the principal empirical finding of our first study, a strong correlation between the {\sl GALEX} FUV-NUV colour and $\Upsilon_*$. We show that 
this finding is not the result of spectral distortions limited to a single passband (eg. metallicity-dependent line-blanketing in the NUV band),  or of the analysis methodology used to measure $\Upsilon_*$,  or of the inclusion or exclusion of the correction for stellar population effects as accounted for using $\Upsilon_*/\Upsilon_{Sal}$.
The sense of the correlation is that galaxies with larger $\Upsilon_*$, or larger $\Upsilon_*/\Upsilon_{Sal}$, are bluer in the UV.  We conjecture that differences in the low mass end of the stellar initial mass function, IMF, are related to the nature of the extreme horizontal branch stars generally responsible for the UV flux in ETGs. If so, then UV color can be used to identify ETGs with particular IMF properties and to estimate $\Upsilon_*$. We also demonstrate that UV colour can be used to decrease the scatter about the Fundamental Plane and Manifold, and to select peculiar galaxies for follow-up with which to further explore the cause of variations in $\Upsilon_*$ and UV colour.
\end{abstract}


\section{Introduction}
\label{sec:intro}

\begin{figure*}
\centering
\includegraphics[scale=0.7]{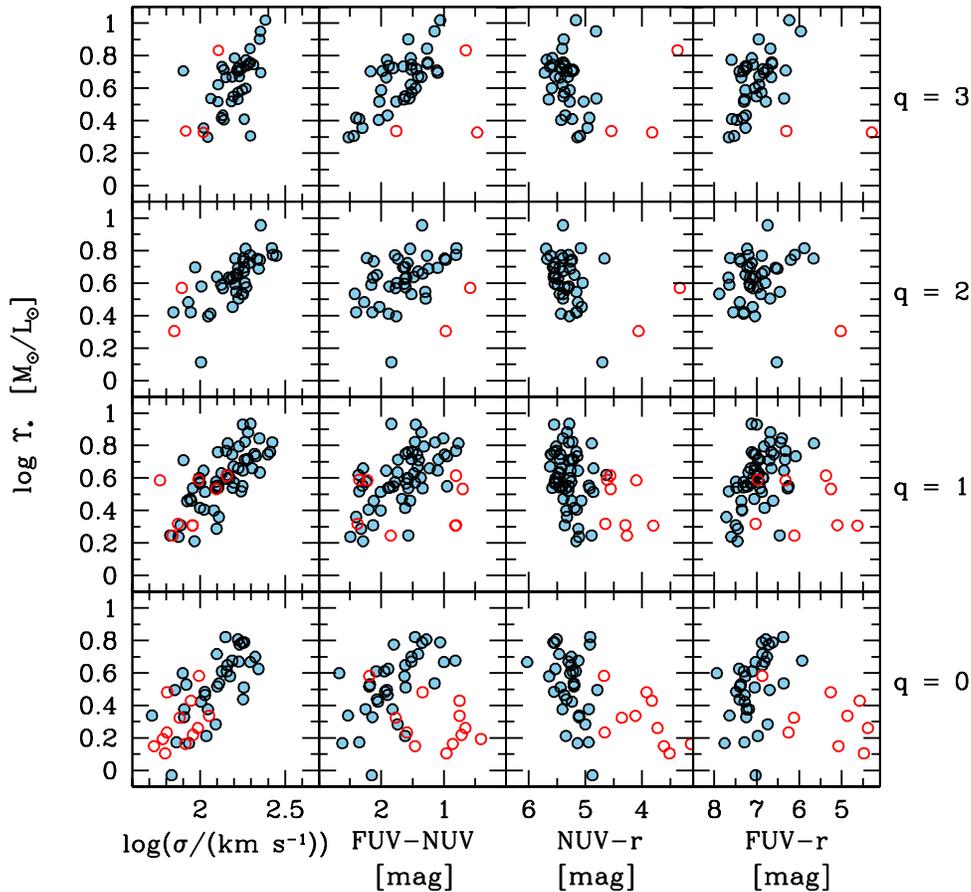}
\caption{Correlation between $\Upsilon_*$ and colours as function of the quality, $q$, of the $\Upsilon_*$ determination (higher $q$ denotes a more reliable determination as
judged by \citet{atlas1}). Galaxies denoted with red circles are removed from further consideration on the basis of the NUV$-$K cut (sources with colour $>$ 7.5 retained) discussed in the text. Colour axes are flipped from standard convention, in our case having blue on the right, so that the $\Upsilon_*$ vs. FUV$-$NUV relation has a slope of the same sign as the $\Upsilon_*$ vs. $\sigma_V$ relation to facilitate comparison.}
\label{fig:qual}
\end{figure*} 

In our first study of this topic \citep[][hereafter Paper I]{zgb}, we identified a correlation between the UV colour of early type galaxies, ETGs, and their stellar mass-to-light ratio, $\Upsilon_*$. If variations in $\Upsilon_*$ represent systemic variations in the stellar initial mass function (IMF) among ETGs, a natural, but not unique, inference is that those stellar population changes are also responsible for the correlated variations in UV colour. As such, we argued that the correlation could provide insight into 
two key unresolved questions regarding the stellar populations of ETGs : the nature of the variations of the IMF and the origin of the unexpectedly high UV flux in many ETGs \citep{code}.

An emerging consensus is that 
the IMFs of ETGs vary systematically with global properties such as velocity dispersion, $\sigma_V$, and metallicity indices such as Mg$_2$  \citep{treu,vandokkum,cappellari,spiniello,ferreras,laesker} and radially within galaxies \citep{martin-navarro}.
The evidence for IMF variations extends beyond ETGs, with stellar clusters providing some of the most direct evidence \citep{strader,z12,z13,z14}. 
Variations in the proportion of low mass stars from system to system, or perhaps in other population characteristics such as the multiplicity function, might manifest themselves in ways beyond mass estimates and star counts.  

The low mass stars invoked by a bottom heavy IMF, when on the main sequence and in the inferred numbers, contribute significantly to the total mass of the stellar population, but not to its luminosity, which is why their relative number alters $\Upsilon_*$. Although, in principle, as these stars evolve they will experience more luminous phases, and so could alter the luminosity, colour, or integrated spectrum of the stellar population, in practice there has been insufficient time for stars with mass $<$ 0.9 M$_\odot$ to evolve off the main sequence. 

If these stars are in binary systems, where one star is sufficiently more massive to have evolved off the main sequence, and if the low mass star plays a role in that evolution, the integrated spectrum of the stellar population may be indirectly affected by the presence of these ``excess" low mass stars. Such modifications of the spectra may be more pronounced at wavelengths where binary stars are relatively most important. At UV wavelengths, where extreme horizontal branch stars dominate the flux of most ETGs \citep{oconnell}, we may find evidence of such a low mass population if, as proposed by \cite{han}, binary stellar evolution plays a role in the origin of EHB stars.

There are alternate hypotheses for the origin of EHB stars that do not involve binaries. These break down into two categories, ``metal-poor" \citep{lee94,park} and ``metal-rich" \citep{bressan,dorman,yi}, and attribute the EHB properties to differences in stellar evolution, driven by metallicity. The motivation for such models comes from the strong correlation between UV upturn strength and metallicity \citep{burstein}. In such models, the correlation of UV properties with $\Upsilon_*$ would be due to a shared dependence of EHB evolution and the IMF on metallicity.  A better understanding of the relationships between Mg$_2$, $\Upsilon_*$, $\sigma_V$, and UV flux and colour among ETGs may help us distinguish among these two broad alternatives and, by doing so, aid in the identification of the origin of both the EHB stars and the IMF variations.

In Paper I we uncovered the correlation between the UV colours and $\Upsilon_*$ using a sample of 32 galaxies, where $\Upsilon_*$ was derived from an analysis of population-sensitive features in the integrated optical spectra \citep{conroy}. In that small sample, we were unable to demonstrated the supremacy of Mg$_2$ or $\Upsilon_*$ in driving FUV$-$NUV.  Furthermore, because the measurements of $\Upsilon_*$ came from spectral synthesis, there was a lingering uncertainty as to whether there exists an internal dependence of $\Upsilon_*$ estimates and Mg$_2$. 
We now examine the relevant properties of 192 galaxies from the ATLAS$^{\rm 3D}$ sample \citep{atlas1,atlas2}. Beside the significant increase in sample size, the other principal difference between the two studies is that the values of $\Upsilon_*$ for this sample come from a dynamical analysis. In \S2  we present measurements from a reanalysis of the FUV and NUV images from $GALEX$ data for this set of galaxies and in \S3 we describe our findings. 

\section{The Data and Measurements}
\label{sec:data}

The parent sample for this study is the set of ATLAS$^{3{\rm D}}$ galaxies with published $\Upsilon_*$ \citep{atlas1,atlas2}. To be specific, we adopt the values of the stellar mass-to-light ratio they derived from subtracting the contribution from a model dark matter halo to their dynamically derived total mass-to-light ratio (fourth column of Table 1 in \cite{atlas2}) and the stellar mass-to-light value referenced to the mass-to-light ratio of a stellar population with a Salpeter IMF and equivalent age and metallicity, $\Upsilon_*/\Upsilon_{Sal}$ (using $\Upsilon_{Sal}$ from the 5th column of the same Table). We track both values because although in principle the referenced value is the more appropriate one, since it aims to remove any dependence on known factors affecting mass-to-light variations, there is some potential for degeneracy in that value because the correction itself depends on modeling the effect of metallicity on stellar evolution.

Our addition to the existing data for this sample is the set of homogeneous measurements of the FUV (1350--1750\AA) and NUV (1750--2750\AA) photometry made possible by the Galaxy Evolution Explorer ($GALEX$) satellite \citep{martin}. We undertake a photometric analysis of galaxies in the ATLAS$^{\rm 3D}$ sample with available
pipeline-processed (GR6/7 release) {\sl GALEX} data from the MAST archive maintained by the Space Telescope Science Institute. There are 231 such targets with
both FUV and NUV data, but for only 199 of these are the images deep enough to enable a proper measurement of the asymptotic magnitudes. Of these, seven are not included in Table 1 of \cite{atlas1} and so are not included here as well, leaving us with a sample of 192 galaxies. The photometry comes from a larger effort to reanalyse nearby galaxy {\sl GALEX} photometry \citep[see][]{bouquin}.

We follow the procedure
described by \cite{gdp} and \cite{lee}. In summary, the analysis steps
are (1) sky-background subtraction,
using elliptical annuli centreed on the galaxy that match the
ellipticity and position angle (PA) of the galaxy and have major axes significantly larger
than the isophotal diameter, D25, in all cases (the same region is used for each of the two UV
bands), (2) interactive masking of foreground stars and background
galaxies following an automated detection of all red (FUV-NUV $>1$) point sources as potential
contaminants, 3) surface photometry within elliptical annuli with
fixed centre, ellipticity and PA (those of the D25 ellipse) and (4)
calculation of the growth curve in both UV bands and the
derivation of the corresponding asymptotic magnitudes.
The asymptotic values are
obtained from a fit to the growth curve, $m(R) \  {\rm vs.} \ dm/dR$,
so they do not include the (systematic to all galaxies)
calibration zero point error (conservatively $\sim$0.1 mag in each band). All surface brightness profiles, colour profiles and
apertures, and asymptotic magnitudes and colours are corrected attenuation adopting
$A_{\rm FUV}=7.9E({\rm B}-{\rm V})$ and $A_{{\rm NUV}} =8.0E({\rm B}-{\rm V})$.
We use the FUV and NUV
asymptotic magnitudes as the best measure of the total UV emission of
our galaxies (Table \ref{tab:dat}).

\section{Results}

We present the UV properties of the sample in Figure \ref{fig:qual} and Table \ref{tab:dat}.  In this first Figure we are addressing two issues. First, there are known cases of ETGs with ongoing, low-levels of star formation \citep[cf.][]{yi,kaviraj}. The emission from these young stars will contaminate our UV fluxes and colours and so we want to remove galaxies in which this component is conspicuous. We adopt a cut in NUV$-$K (fsources with colour $>$7.5 retained) to select galaxies on the red sequence that also have relatively low molecular hydrogen,  H$_2$ \citep[Figure 8 from][]{young}. The unwanted objects are shown in the panels with red open circles and are removed from subsequent discussion. 
Second, the dynamical measurements of the {\sl total} mass-to-light ratio derived by \cite{atlas1},  $\Upsilon_{\rm JAM}$, come with quality flags, which we propagate to the measurements of the {\sl stellar} mass-to-light ratios, $\Upsilon_*$, presented in \cite{atlas2}. While we would prefer to use all of the data for statistical reasons, we do not want to dilute our analysis with unreliable values of $\Upsilon$. Therefore, in the Figure we separate the galaxies for the four different quality levels (3,2,1,0, with 3 being best). After the removal of the sources we believe have residual star formation (sources with NUV$-$ K $\le$ 7.5 removed), the correlations (or lack thereof) do not appear to change markedly among the different quality determinations. Therefore, we make no additional cut based on the provided quality diagnostic and are able to retain as large a sample as possible.

\begin{figure}
\centering
\includegraphics[scale=0.4]{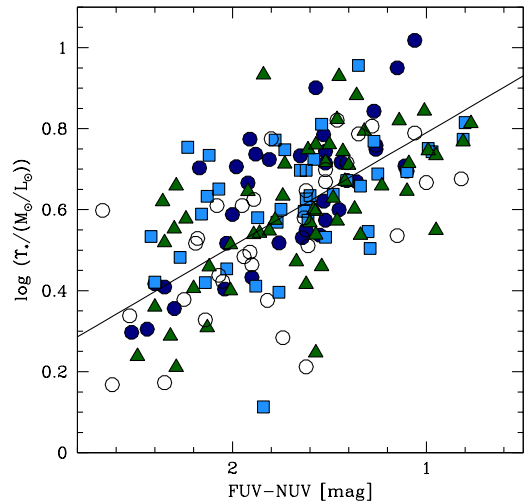}
\caption{$\Upsilon_*$ vs. UV colour. Points are coded on the basis of the $\Upsilon_{\rm JAM}$ quality flag (in order of decreasing quality: filled circles, filled squares, filled triangles, and open circles). The orthogonal regression fit \citep{isobe} is shown ($\log \Upsilon_* = 1.07 - 0.28 ({\rm FUV} - \rm{NUV})$). The colour axis remains inverted from normal convention for consistency with Figure \ref{fig:qual}. UV blue galaxies are on the right. The Spearman rank correlation test indicates that the probability that the observed correlation arose by chance is $4.4 \times 10^{-13}$.}\label{fig:uvcol}
\end{figure} 

In Figure 1, the correlation between FUV$-$NUV and $\Upsilon_*$ is visually indistinguishable from that between $\sigma_V$ and $\Upsilon_*$, even though $\sigma_V$ is a key parameter in the determination of $\Upsilon_*$. Therefore, we are already assured that the principal result from Paper I, that the correlation between $\Upsilon_*$ and UV colour is as strong as any other found so far with $\Upsilon_*$, is confirmed.

We combine the different quality subsets and present the composite correlation between FUV$-$NUV and $\Upsilon_*$ in Figure \ref{fig:uvcol}. Using the Spearman rank correlation coefficient, we calculate that the chance of randomly realizing such a correlation is $4.4 \times 10^{-13}$, and is 7.6$\times10^{-5}$ for $\Upsilon_*/\Upsilon_{Sal}$, so highly significant in either case, albeit weaker in the latter. In the Figure we also present a fitted line (using the orthogonal regression formula from \cite{isobe}) and find that $\log \Upsilon_* = 1.07 - 0.28({\rm FUV}-{\rm NUV})$.

\begin{figure*}
\centering
\includegraphics[scale=0.7]{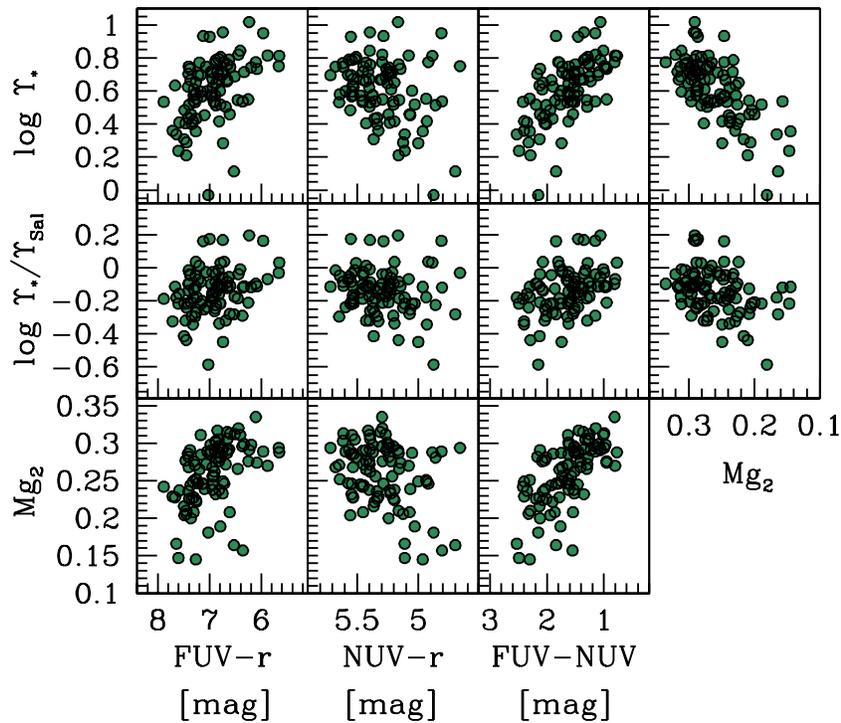}
\caption{Correlations between $\Upsilon_*$, FUV$-$NUV, FUV$-$r, NUV$-$r, and Mg$_2$. $\Upsilon_*$ is in solar units. 
We use only the subsample of galaxies for which Mg$_2$ is available from \citet{golev}. Colour axes remain reversed from convention for consistency throughout this paper.}
\label{fig:mg2}
\end{figure*}

Next, we compare the correlations between $\Upsilon_*$ and other quantities. In particular, the Mg$_2$ index has been found to correlate strongly with $\Upsilon_*$ \citep{conroy} and with UV colour \citep{burstein,donas}. We adopt measurements of Mg$_2$ presented by \cite{golev}, although unfortunately measurements only exist for a subsample of our galaxies.
Thus, we drop from the 166 galaxies included in Figure \ref{fig:uvcol} to the 93 presented in Figure \ref{fig:mg2}.
We calculate the various Spearman correlation coefficients and the associated probabilities that they arise by chance (Table \ref{tab:spear}) for the subset of our 93 galaxies with published Mg$_2$ from \cite{golev}. In all cases where a correlation is detected and Mg$_2$ is not required, and so cases where we can use more galaxies, the correlation is found to be even more statistically significant than quoted in the Table when we use all of the available data.

There are various results to discuss from Figure \ref{fig:mg2} and Table \ref{tab:spear}. First, we confirm the finding from Paper I that the correlation between UV colour and $\Upsilon_*$ is as strong, and perhaps stronger, than that with Mg$_2$. We extend that result and find that it is also valid using the corrected values of the mass-to-light ratio, $\Upsilon_*/\Upsilon_{Sal}$. As noted in our first paper, one cannot read too much into a comparison of the Spearman rank correlations (or associated probabilities) because the observational scatter is different among different observables. However, there is no doubt that all correlations between UV colour, $\Upsilon_*$, and $\Upsilon_*/\Upsilon_{Sal}$ are highly significant. 

Second, there is a reversal of the prevailing trends when examining NUV$-$r, in the sense that larger $\Upsilon_*$ values correspond to redder galaxies. \cite{donas} identified the correlation between NUV$-$V and Mg$_2$ and suspected it arose due to increased line-blanketing in the NUV passband in the higher metallicity galaxies. The correlation we find between NUV and r goes in that same sense, namely that as Mg$_2$ increases, NUV decreases (assuming r is fixed). This correlation is weaker than some of the other correlations and only marginally significant for this sample (probability of arising by chance = 0.015), but is highly significant for our entire sample (probability of arising by chance $= 5.8 \times 10^{-6}$). If the cause of this correlation is indeed blanketing in the NUV, then blanketing will also affect FUV$-$NUV in the sense that as Mg$_2$ increases, FUV$-$NUV will decrease as the NUV flux becomes increasingly diminished (if FUV is fixed). As such, this would appear to be a consistent explanation of the correlations we find between FUV$-$NUV vs. Mg$_2$ and NUV vs. r. 

However, we see from the panels showing FUV$-$r vs. $\Upsilon_*$ and FUV$-$r vs. Mg$_2$, and Table \ref{tab:spear}, that the trends of UV colour with $\Upsilon_*$ and Mg$_2$ are present even when not using NUV, and that they are much more significant than those involving NUV (the probability of FUV$-$r and NUV$-$r vs. $\Upsilon_*$ arising by chance are $2.7 \times10^{-6}$ vs 0.015, respectively). Therefore, the observed FUV$-$NUV  vs. $\Upsilon_*$ trend does not arise primarily from blanketing of the NUV flux, although such blanketing will help strengthen the UV color relation by decreasing the NUV flux as the FUV flux increases. Because the qualitative nature of the correlations remains in place whether one uses FUV or FUV$-$NUV and $r$ or K (the latter not shown, but the trends are nearly indistinguishable to what is seen in r), we conclude that the results are not driven by the ``artificial" suppression of flux in any one passband.

A critical question is whether the trend we find between UV colour and $\Upsilon_*$ is merely a reflection of the previously identified correlations between UV colour and Mg$_2$ \citep{burstein} and $\Upsilon_*$ and Mg$_2$ \citep{conroy}, or something more fundamental. This is always a difficult class of question to answer in practice because of the interrelation among parameters and unequal uncertainties for the different parameters. One way to begin to address this issue is to remove the known influence of metallicity on $\Upsilon_*$. We have previously alluded to this approach in the use of $\Upsilon_*/\Upsilon_{Sal}$. When we use $\Upsilon_*/\Upsilon_{Sal}$, we remove the effect of metallicity on the observed mass-to-light ratio of a stellar population because models account for the role of metallicity in the luminosity and colors of the stellar population. The choice of a Salpeter IMF is incidental, we are simply referencing the observed values to model values. We find that the correlation between UV colour and $\Upsilon_*/\Upsilon_{Sal}$ is highly significant (Table 2), demonstrating that although Mg$_2$ does influence the UV colour of the stellar populations, it is not the sole factor. The strength of this method is that we are applying what we already know to be the case, that abundances affect the colors of stars. The weaknesses include the assumption that the modeling is correct (one could in principle introduce a correlation with the wrong model) and that we are only correcting for this one possible role of abundance.

For an alternative approach we use
partial rank correlation coefficients \citep{kendall}. 
 The expression for the partial correlation coefficient is
$$\rho_{x,y;z} = (\rho_{x,y}-\rho_{x,z}\rho_{y,x})/((1-\rho^2_{x,z})(1-\rho^2_{y,z}))^{1/2},$$
where $\rho_{x,y}$ is the Spearman rank correlation coefficient for variables $x$ and $y$. 
The partial correlation coefficient measures the rank correlation, $\rho$, between two variables (to the left of the semicolon in the $\rho$ subscript), where the effect on the correlation of the variable to right of semicolon is removed.  This approach is aimed at removing any possible influence that abundance, as measured by Mg$_2$, has on the correlation between colour and the mass-to-light ratio. Using the 93 galaxies in the sample for which we have Mg$_2$ measurements, we find that $\rho_{UV,\Upsilon_*;Mg_2} = -0.391$. The agreement between the correlation coefficient obtained when we remove the known effects of abundance through modeling ($-0.399$) and through the model-free partial correlation coefficient analysis $(-0.391)$, suggests that the bulk of the influence of abundance is through the colour of the stellar population and provides supporting evidence for a direct correlation between the IMF and UV colour.

\subsection{Putting the correlation to use}

As with any correlation among galaxy parameters, there are two paths forward. To provide a specific example, consider the relationship between luminosity and H{\small I} line-width, generally referred to as the Tully-Fisher relation \citep{tully}. Clearly, we desire to understand the physics behind that relation. On the other hand, we use the relation, without understanding the physics, as a distance estimator. Likewise here, we need to understand the cause of the FUV$-$NUV vs. $\Upsilon_*$ vs. Mg$_2$ relationships, but we can also attempt to exploit them in other causes.

In Figure \ref{fig:uvcol} we show the relationship between FUV$-$NUV and $\Upsilon_*$. One can use this relationship, and a measurement of FUV$-$NUV, to obtain rough estimates of $\Upsilon_*$ for much larger samples of early-type galaxies than one could hope to do with either the dynamical approach of \cite{cappellari} or the spectral method of \cite{conroy}.

\begin{figure}
\centering
\includegraphics[scale=0.4]{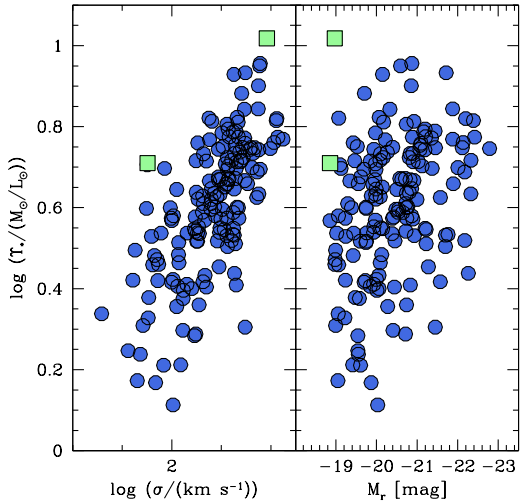}
\caption{Using FUV$-$NUV to select interesting subpopulations for detailed studies. We illustrate that one possible set of criteria (FUV$-$NUV $<$1.5 and M$_r < 19$) selects a set of low-luminosity, high $\Upsilon_*$ systems that might prove useful for detailed follow-up studies to explore the origin of high $\Upsilon_*$. The two interesting objects are highlighted as squares.}
\label{fig:selection}
\end{figure}

\begin{figure}
\centering
\includegraphics[scale=0.4]{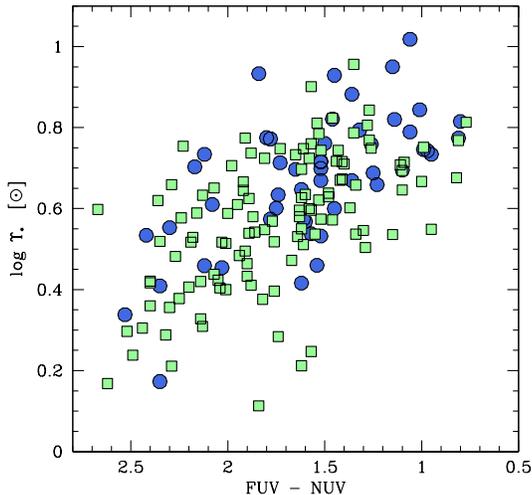}
\caption{Type dependence of $\Upsilon_*$. We divide the sample at a T-type of $-$3, with galaxies of T-type $\ge -3$ being S0's and later (green squares), while those with T-type $<-3$ are ellipticals (blue circles). There is no evident difference in the relation among the two classes of galaxy.}
\label{fig:type}
\end{figure}

As an example of how this relation could be exploited to select physically interesting galaxies from a large sample, 
we select from our sample those galaxies with FUV$-$NUV $< 1.5$ and M$_r < 19$. The resulting two galaxies are highlighted in Figure \ref{fig:selection}. With this photometric selection we have identified two of the most distinct galaxies in the ATLAS$^{\rm 3D}$ sample. NGC 4342 is the highlighted galaxy at the top of the panels, with the largest $\Upsilon_*$. This is an S0. The other galaxy is NGC 525, also an S0, that is apparently fairly extreme in having a velocity dispersion $<$ 100 km s$^{-1}$. Because it is an S0, there may be significant rotation and although it is intrinsically faint, it has a fairly large $\Upsilon_*$. That both of these galaxies are S0's may cause some concern, but we show in Figure \ref{fig:type} that dividing the sample at a T-type of $-3$ demonstrates that there are no evident systematic differences in $\Upsilon_*$ between ellipticals and lenticulars. Photometrically selecting such galaxies from large samples for detailed follow-up, such as the 2-D spectroscopy necessary for the sophisticated dynamical analysis of \cite{atlas1}, will enable investigators to tune their samples to best address questions regarding the origins of the IMF or UV physics.

\begin{figure}
\centering
\includegraphics[scale=0.4]{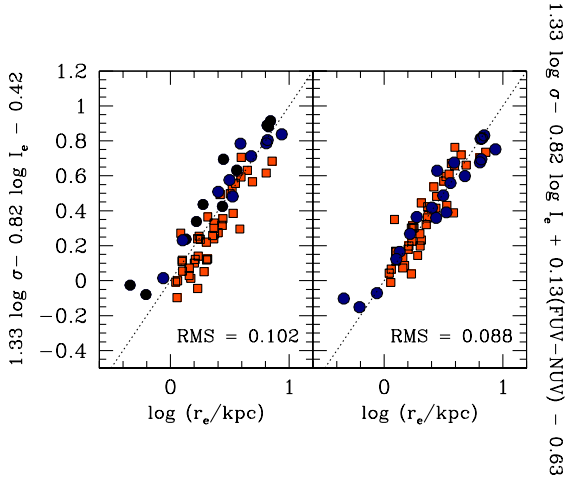}
\caption{Fundamental Plane residuals and UV colour. We only show the extremes of the galaxy distribution, galaxies with FUV$-$NUV $> 2.0$ as red squares and those with FUV$-$NUV $< 1.2$ as blue circles for clarity. In the left panel is the original FP relation. Systematic differences based on UV colour are visible. We fit the relationship between residual and UV colour and correct for that in the right panel. The RMS scatter about each FP is given in the associated panel and is the result for the full sample. The velocity dispersion $\sigma$ is in units of km s$^{-1}$ and the mean surface brightness interior to r$_e$, I$_e$, is in units of L$_\odot$ pc$^{-2}$. }
\label{fig:fp}
\end{figure}

The correlations found here can also be used to refine other known correlations. As an example, we examine the Fundamental Plane (FP) relation of early types. In Figure \ref{fig:fp} we plot the FP found by \cite{bernardi}, $\log r_e = 1.33 \log \sigma_V - 0.82 \log I_e + C$ in the left panel of the Figure, where $r_e$ is the effective radius and $I_e$ is the surface brightness within $r_e$, all quantities for these galaxies coming from \cite{atlas1}. We have added a constant term to have the data pass through the 1:1 line. The data are colour coded by FUV$-$NUV colour and there is an evident preference for the redder points to lie below the mean relation and bluer points above (these colour match the sense of the FUV$-$NUV colours). Fitting a relationship for the residual vs. FUV$-$NUV colour, we arrive at the FP on the right, which has a lower rms scatter (the colour sensitive correction involves subtracting $0.21 - 0.13({\rm FUV}-{\rm NUV})$. This exercise demonstrates that FUV$-$NUV can be used as an additional parameter with which to refine the FP, presumably because variations in $\Upsilon_*$ generate scatter about the FP and the UV colour provides information on $\Upsilon_*$. 

Various refinements and extensions of the FP have been presented \citep{zfm1,zfm2,mcgaugh,dutton,atlas1} where the full dynamical support is included, rotation and pressure support, the curvature of the relation between the total $\Upsilon$ and $\sigma_V$ is accounted for, and attention is paid to expressing the relation in terms as closely related to the Virial theorem as possible to ease interpretation. With these changes the relationship is more accurate for very low and high mass ellipticals and is connected to later type galaxies as well. At the heart of these efforts is the recovery of the mass enclosed within the radius of interest, typically the effective radius $r_e$. In the case of \cite{atlas1}, they construct the ``mass plane"  relation because they have the data required to measure a dynamical $\Upsilon$. On the other hand, if one does not have access to such data (as would be the case for much larger samples) then perhaps one is limited to a fitting function for $\Upsilon$ of the type presented by \cite{zfm2}. In that case, refinement of those estimates obtained by employing  FUV$-$NUV could prove useful. 

\begin{figure}
\includegraphics[scale=0.4]{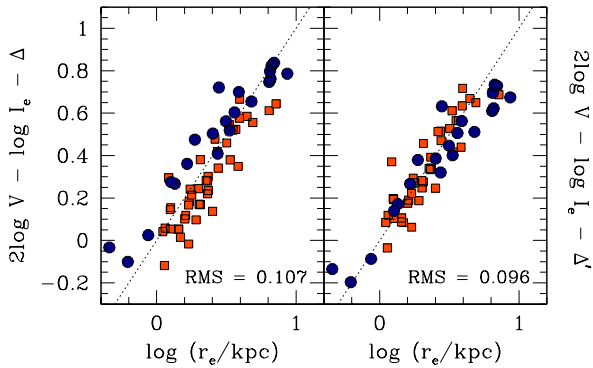}
\caption{Fundamental Manifold residuals and UV colour. We divided the sample as in Figure \ref{fig:fp} by UV colour and now plot the FM \citep{zfm2} in the left panel. Again, systematic differences based on UV colour are visible. Those are removed in the right panel, with the resulting lower RMS. The units of V are km s$^{-1}$ and of I$_e$ are L$_\odot$ pc$^{-2}$.}
\label{fig:fm}
\end{figure}

When we plot the galaxies on the Fundamental Manifold \citep{zfm1,zfm2} in Figure \ref{fig:fm}, we find a similar result as for the FP in that the residuals correlate with UV colour and that correcting for that trend results in a lower RMS version of the plot (colour sensitive correction converts $\Delta + 0.25 - 0.143({\rm FUV}-{\rm NUV})$ to $\Delta^\prime$). For technical reasons, the comparison with the FP is interesting. The FM uses a functional fit to model the behavior of $\Upsilon$'s dependence on the kinematic term $V$, $V \equiv \sqrt{v_{rot}^2/\alpha + \sigma_V^2}$, where $\alpha$ is typically about 2 and $v_{rot}$ is a measure of the bulk rotational velocity, and $I_e$. Fitting this function would have removed as much of the dependence of $\Upsilon_*$ on either $V$ or $I_e$ as possible. Our finding that FUV$-$NUV can be used to decrease the scatter of the FM means that there are variations in $\Upsilon_*$ that are not correlated with either of those terms primarily (or sufficiently broadly to have been fitted out). In other words, variations in $\Upsilon$ arising from an IMF that is determined by $\sigma$ would have been fitted out in the original determination of $\Delta$. 

We close by noting that far more detailed work has been presented on uncovering differences among stellar populations of galaxies within the FP. Both \cite{graves} and \cite{springob} present evidence for systematic variations that contribute to the thickness of the edge-on projection of the FP. This work does not supplant those studies, but instead offers an  observationally efficient way to reduce that thickness using a measurement, the UV colour, that is simpler to obtain than the detailed line ratios used in those studies. Of course, the detailed line ratios and subsequent stellar population modeling provide a far more in depth view of the stellar populations than does the UV colour, and by doing so help address questions regarding the root origin of the scaling relations, but that is not our focus here.

An avenue that appears to have significant promise is the use of spatially resolved IMF variations, abundances, and UV colours within individual galaxies. Comparing to data published by \cite{martin-navarro}, we find than in the three galaxies in common between our samples there are qualitatively corresponding gradients in the IMF variations as they determined through spectral line analysis and our UV colours.  Galaxies NGC 4552 and NGC 5557 both exhibit a gradient in the inferred IMF within r$_e$ and a corresponding FUV$-$NUV colour gradient in our data, while NGC 4387 shows no gradient in either measurement. In combination with abundance measurements and detailed stellar population modeling, such cases may provide a path forward in this research area.

\section{Conclusions}

Using {\sl GALEX} data in combination with published values of the stellar mass-to-light ratio, $\Upsilon_*$ for a large sample of early type galaxies \citep{cappellari}, we have confirmed our previous identification of a strong correlation between UV colour (FUV$-$NUV) and $\Upsilon_*$ in early type galaxies (ETGs). The correlation is also highly significant between UV colour and $\Upsilon_*/\Upsilon_{Sal}$, demonstrating that the results reflect more than just known differences among stellar populations. The key differences between this study and that presented in Paper I are that we use a much larger sample, that the values of $\Upsilon_*$ were obtained in a completely independent manner (dynamical analysis vs. spectral line modeling), and that we remove the role of metallicity (alpha enhancement) on the results both through modeled corrections (the use of $\Upsilon_{Sal}$) and more detailed statistics (the use of partial correlation coefficients) . 

If variations in the stellar initial mass function are indeed responsible for the variations in $\Upsilon_*$, then they may also be responsible for the variations in UV colour. This conjecture provides an avenue to addressing the long-standing question regarding the origin of the UV flux in ETGs and identifies a photometric signature of IMF variations. 

The existence of the correlation between UV colour and $\Upsilon_*$  (and $\Upsilon_*/\Upsilon_{Sal}$) is something we need to understand, but we can also use it to aid us in other programs. We can use our findings to easily select objects that are likely to have unusual values of $\Upsilon_*$ for further study, or to help reduce the scatter in galaxy scaling relations that depend on $\Upsilon_*$. With the advent of homogeneous, precise UV magnitudes for large samples \citep{bouquin}, it is now time to use the correlations described here to revisit the stellar populations of statistical samples of early type galaxies.

\section{acknowledgments}

DZ acknowledges financial support from 
NASA ADAP NNX12AE27G and  NYU CCPP for its hospitality during long-term visits. 
The authors acknowledge the support from the FP7 Marie Curie Actions of the European Commission, via the Initial Training Network DAGAL under REA grant agreement PITN-GA-2011-289313.
This research has made use of the NASA/IPAC Extragalactic Database (NED), which is operated by the Jet Propulsion Laboratory, California Institute of Technology, under contract with NASA.

\begin{table}
\caption{Galaxy Sample and Properties$^{a,b}$}
\begin{tabular}{lrrr}
\hline
Name & $m_{FUV}$ & $m_{NUV}$ & Distance \\
&&&[Mpc]\\
\hline
IC0676&17.85$\pm$0.02&16.51$\pm$0.02&24.6\\
IC0719&16.91$\pm$0.01&16.26$\pm$0.01&29.4\\
IC0782&20.50$\pm$0.07&18.12$\pm$0.02&36.3\\
IC1024&16.94$\pm$0.02&16.36$\pm$0.02&24.2\\
IC3631&17.35$\pm$0.04&16.55$\pm$0.02&42.0\\
NGC0448&19.38$\pm$0.33&17.62$\pm$0.04&29.5\\
NGC0474&18.44$\pm$0.17&16.08$\pm$0.05&30.9\\
NGC0509&20.25$\pm$0.31&18.34$\pm$0.06&32.3\\
\hline
\end{tabular}
\break
{$^a$}{The first few entries of the table are shown here for guidance. The complete table is available on-line.}
\break
{$^b$}{The quoted uncertainties do not include the zero point uncertainties, but a systematic zero  point error would affect all measurements equally and  so do not affect the conclusions presented here.}
%\enddata
\label{tab:dat}
\end{table}

\begin{table*}
\caption{Rank correlation coefficients and probabilities that they arise randomly$^a$}
\begin{tabular}{lrrrrrrrr}
\hline
&FUV $-$ NUV&FUV $-$ r&NUV $-$ r&$L_r$&$\sigma_V$&Mg$_2$&$\Upsilon_*$&$\Upsilon_*/\Upsilon_{Sal}$\\
\hline 
FUV $-$ NUV&...&8.2$\times10^{-29}$&0.084&0.079&3.3$\times10^{-10}$&2.7$\times 10^{-14}$&4.4$\times 10^{-13}$&7.6$\times10^{-5}$\\
FUV $-$ r      &{\bf 0.864}&...&0.008&0.397&2.2$\times10^{-5}$&4.8$\times 10^{-8}$&2.7$\times 10^{-7}$&0.0054\\
NUV $-$ r      &$-$0.180&0.274&...&0.123&0.012&0.015&0.015&0.131\\
L$_r$             &0.183&0.089&$-$0.161&...&1.1$\times 10^{-7}$&0.001&0.061&0.965\\
$\sigma_V$        &{\bf $-$0.595}&{\bf $-$0.425}&0.258&{\bf $-$0.517}&...&6.6$\times 10^{-15}$&1.9$\times 10^{-17}$&8.2$\times10^{-9}$\\
Mg$_2$              &{\bf $-$0.688}&{\bf $-$0.530}&0.253&$-$0.326&{\bf 0.699}&...&1.7$\times 10^{-12}$&0.0005\\
$\Upsilon*$    &{\bf $-$0.663}&{\bf $-$0.503}&0.252&$-$0.195&{\bf 0.742}&{\bf 0.651}&...&9.1$\times10^{-30}$\\
$\Upsilon_*/\Upsilon_{Sal}$&{\bf $-$0.399}&{\bf $-$0.286}&0.158&$-$0.005&{\bf 0.554}&{\bf 0.354}&{\bf 0.871}&...\\
\hline
\end{tabular}
\label{tab:spear}
\bigskip
\break

{$^a$}{Rank correlation coefficients given in bottom half of matrix, probabilities in top half. Significant correlations (probability of arising by chance $< 0.01$) highlighted in bold.}
\end{table*}

\bibliography{biblicomplete}
\end{document}